\documentclass[nofootinbib,twocolumn,
showpacs,preprintnumbers,amsmath,amssymb]{revtex4}
\usepackage{latexsym}
\usepackage{hyperref}
\usepackage{epsfig, psfrag, graphicx}

\newcommand{\be}{\begin{equation}}
\newcommand{\ee}{\end{equation}}
\newcommand{\ben}{\begin{eqnarray*}}
\newcommand{\een}{\end{eqnarray*}}
\newcommand{\bea}{\begin{eqnarray}}
\newcommand{\eea}{\end{eqnarray}}
\newcommand{\bdm}{\begin{displaymath}}
\newcommand{\edm}{\end{displaymath}}
\newcommand{\ba}{\begin{align}}
\newcommand{\ea}{\end{align}}
\newcommand{\lb}{\label}
\renewcommand{\d}{\operatorname{d}\!}
\renewcommand{\exp}{\operatorname{exp}\!}
\renewcommand{\cosh}{\operatorname{cosh}\!}
\renewcommand{\sinh}{\operatorname{sinh}\!}

\renewcommand{\ln}{\operatorname{ln}\!}
\newcommand{\del}{\partial}

\begin{document}

\title{\bf Quantum cosmology with big-brake singularity}

\author{Alexander Y. Kamenshchik}
\affiliation{Dipartimento di Fisica and INFN, Via 
Irnerio 46, 40126 Bologna, Italy\\
L.D. Landau Institute for Theoretical Physics of the 
Russian Academy of Sciences, Kosygin str. 2, 
119334 Moscow, Russia}
\author{Claus Kiefer}
\affiliation{Institut f\"ur Theoretische Physik, Universit\"{a}t zu K\"{o}ln, Z\"{u}lpicher
Strasse 77,
50937 K\"{o}ln, Germany}
\author{Barbara Sandh\"ofer}
\affiliation{Institut f\"ur Theoretische Physik, Universit\"{a}t zu K\"{o}ln, Z\"{u}lpicher
Strasse 77,
50937 K\"{o}ln, Germany}
\begin{abstract}
We investigate a cosmological model with a big-brake singularity
in the future:
while the first time derivative of the 
scale factor goes to zero, its second time derivative tends to minus
infinity. Although we also discuss the classical version of the model
 in some detail, our main interest lies in
its quantization. We formulate the Wheeler--DeWitt
equation and derive solutions describing wave packets. We show that
all such solutions vanish in the region of the classical singularity,
a behaviour which we
interpret as singularity avoidance. We then discuss the same
situation in loop quantum cosmology. While this leads to a different
factor ordering, the singularity is there avoided, too. 
\end{abstract}

\maketitle

\section{Introduction}
It is a well-known fact that general relativity is an incomplete
theory in the sense that solutions to Einstein's equations can
contain singularities. These are regions (outside spacetime) where the
theory itself breaks down. According to the 
singularity theorems, the occurrence of such singularities is a generic
feature of `physical' solutions to Einstein's equations.

One outlook on this problem is to consider a quantum theory of gravity
as the necessary completion of general relativity \cite{OUP}.
 Consequently, it is expected that
such a quantum theory of gravity is in some sense (still to be
specified) free of these singularities. Investigations to this end
are usually carried out, not in the full quantum gravity candidate
theories, but in reduced models. That is, one takes a specific solution
(conventionally and pragmatically specified by some symmetry) to
Einstein's equations and in some way applies the quantization procedure
of the full theory to the reduced model.\footnote{A counter-example is
  causal dynamical triangulation \cite{triangulation,OUP}.
 Here exists the possibility to reduce the full quantum theory by
 integrating out all degrees of freedom except the scale factor. The
 resulting theory yields an action differing from the
 geometrodynamical minisuperspace action by an overall minus sign in
 the realm where the continuum limit is valid. Numerical evaluation
 predicts a closed universe undergoing a bounce upon reaching small
 scales. Moreover, quantum spacetime on these scales is predicted to
 be of fractal structure and dimension $2$, coinciding with results
 obtained in the asymptotic-safety approach \cite{safety,OUP}.}
Prototypes for such symmetry-reduced models are black-hole spacetimes
and cosmological spacetimes. 

In our paper we restrict the discussion to cosmological
models. Here, in the canonical approach, we have basically two candidates for a quantum
cosmological theory:
minisuperspace quantization in the framework of the geometrodynamical 
approach and loop quantum cosmology \cite{OUP,Coule,Bojoreview}.
In both approaches, one has to investigate whether singularities `do not
occur'. This implies that for each approach one has defined what the
sentence `singularities do not occur' means. To come to the point, for
neither of the two theories a strict proof of the avoidance of
singularities exists.

Both approaches describe the universe via a
wave function on configuration space which has to be the solution of
a constraint equation. The constraint equation is the quantized
version of the Hamiltonian constraint. The difference between both
approaches lies in the way this equation is quantized. In loop quantum cosmology, one uses a so-called polymer representation instead of the conventional Schr\"odinger representation. This is done in analogy to the full theory.
This procedure carried out in a naive way, leads to a difference equation in steps of a smallest length $\mu_0$. 
In geometrodynamics, one arrives at a differential equation, the
Wheeler--DeWitt equation. In the continuum limit, $\mu_0\to 0$
(suitable conditions on the higher derivatives of the wave function
implied), the loop quantum cosmological 
difference equation fades into the Wheeler--DeWitt equation
\cite{Bojowald}.  

Recently, Ashtekar {\em et al.} \cite{Ashtekar,Vandersloot}
 extended the ansatz using $\mu_0$,
replacing it by  $\bar{\mu}$, which is a function of the densitized 
triad operator $\hat p$. The equation is then a difference
equation in eigenvalues $v$
 of the volume operator, and the Wheeler--DeWitt
equation follows in the continuum limit for large volume. The
factor-ordering of the Wheeler--DeWitt equation then does depend on
the factor-ordering chosen for the difference equation. In
\cite{Bojowald} and \cite{Ashtekar,Vandersloot} different
factor-orderings have been chosen.\\ 
The two difference equations, in $\mu_0$ or $\bar\mu$, can be
understood in a broader context as implementing different actions of
the full Hamiltonian constraint. They are thus just two special cases
of a wider class of constraints that might arise, the actual form of
which should in principle be determined by the full Hamiltonian
constraint, \cite{LatticeRefinement}. Whereas in the first case, the
coordinate edge length of a holonomy is fixed and does not depend on
the scale factor, in the second case it does. This can be interpreted
as an implementation of the fact that in the full theory, the
Hamiltonian constraint (whatever its exact form may be) creates
vertices (in addition to changing the edge labels of the existing
edges). As new vertices are created, the edge lengths decrease. The
altered dynamics using $\bar\mu$ then corresponds to a lattice in
which the number of vertices grows linearly with volume. 

In loop quantum cosmology, results on singularity resolution fall into
one of three categories, \cite{QuantumSingularity}.  
As a first result one may quote that, in the isotropic case, the
evolution equation is well-defined also on an evolution across the
singularity. This is due to the discreteness of  the evolution
parameter which is a feature inherited from the full theory through
the use of the polymer representation, \cite{Husain}. This allows to
evolve a wave packet, starting from any initial state,
deterministically across the singularity,
\cite{SingularityAvoidance}. \\ 
A second hint on singularity avoidance, so far studied in isotropic
models with massless scalar field $\phi$, curvature index 
${\mathcal K}=0,1$ and
zero as well as non-zero cosmological constant, is the occurrence of a
so-called `bounce'. As a bounce one describes a deviation from the
classical behaviour such that a semi-classical wave packet starting on
a classical trajectory for large scale-factor deviates from this
trajectory upon approach of the classical singularity and instead
avoids the region of configuration space where the singularity is
located. Here, avoidance refers to an exponential fall-off (in $\phi$)
of the wave function, \cite{Bojowald,Ashtekar,Vandersloot}.\\  
A third criterium is given by the boundedness of the expectation value of the operator corresponding to the inverse scale factor. 
As the inverse scale factor is related to the curvature in isotropic,
homogeneous models, this hints at avoidance of the curvature
singularity. 
This is a feature which follows from the use of holonomies as basic
variables. It is a purely kinematical result as the expectation value
is evaluated  with respect to states from the kinematical Hilbert
space, \cite{SingularityAvoidance,Boundedness}.\\  
The robustness of these results is disputable to differing
degree. Whereas the possibility to evolve the wave packet through
singularities in a well-defined way seems to persist in the full
theory, this is not so clear for the other two criteria.\\ 
The boundedness of the inverse scale factor seems to carry over to the
full theory only when evaluated on a subspace of the kinematical
Hilbert space, \cite{Thiemann}. 
Moreover, the occurrence of a bounce seems to be knit to isotropic
models and even there it is not clear whether it should persist for
more general settings involving a matter potential. 
The underlying concept in the models studied in this context
is to use the scalar field as a `time' variable (emergent
time) with respect to which the wave packet is evolved
(numerically). Transferring this concept to more general models
including a scalar-field potential, one has to cope with a `time'
(i.e. $\phi$-) dependent evolution operator which is given by the
square-root of the gravitational Hamiltonian plus the scalar field 
potential energy. It can therefore be arbitrarily complicated. In
addition to that, it is not clear that $\phi$ defines a `good clock'
throughout the universe evolution.  
The advantage of this approach, on the other hand, is 
the existence of an inner product which is uniquely
defined by a complete set of Dirac observables, and thus provides expectation
values of observables. Most importantly, the inner product supplies
the model with a probability interpretation (even though no connection
to the measurement process is made).

In the geometrodynamical framework, several models have been
investigated regarding their ability to resolve the singularity
problem. In this setting, singularity avoidance is defined as either a
vanishing of the wave function at the point of the classical
singularity\footnote{More generally, it would be sufficient to demand
  that the probability vanishes there; for example, the ground-state
  wave function for the hydrogen atom, as found as a solution to the
  Dirac equation, diverges for $r\to0$, but the probability there is
  zero because of the $r^2$-contribution from the measure. In quantum
  cosmology, this question is more subtle because the fundamental
  measure is not known \cite{OUP}.} 
or a spreading of  semi-classical states denoting a
break-down of semi-classical concepts in general (the end of the world
as we know it). In the semi-classical regime, an approximate
Schr\"odinger equation can be derived from a WKB-expansion defining a
notion of time \cite{OUP}. This time label is necessary to stack
together the
$3$-hypersurfaces on which the wave function has support.
The thus obtained $4$-dimensional spacetime can now be
probed for geodesic completeness. Only in semi-classical regimes a
notion of geodesics exists, and thus we can speak of the existence of
singularities --- in the strict mathematical sense of the singularity theorems --- only there.

Accepting both
criteria, singularity avoidance was found for big-bang/big-crunch
singularities in various models (different scalar field potentials,
cosmological constant, etc.) and for the big-rip singularity occuring
at large scale factor, \cite{MCB}. 
The big-rip singularity is a singularity which the universe can encounter when
it expands ``too rapidly'' \cite{star-sing}. This singularity occurs
when the cosmological radius of the universe $a(t)$ tends to infinity
at some finite moment of time simultaneously with its time derivative
$\dot{a}(t)$ in such a way that the Hubble variable 
$H(t) \equiv \dot{a}/{a}$ tends to infinity as well.
Interest in this type of singularity is connected with the fact that
it arises quite naturally in cosmological models with
phantom dark energy,
that is, dark energy such that the equation of state parameter $w =
p/\rho<-1$ \cite{rip,phantom}, where $p$ and $\rho$ denote pressure and
energy density of the cosmological fluid, respectively.

In the following, we want to analyze whether the so-called {\em big-brake
singularity} can be avoided in a similar way.
The big brake belongs to another class of cosmological singularities 
not connected with the divergence of the Hubble variable itself
but of one of its higher derivatives. Singularities of this type are called
soft, quiescent, or sudden \cite{shtanov,Barrow,we-tach}. These
singularities occur at finite value of the scale factor and its time 
derivative and hence of the Hubble parameter, 
while the first or higher derivatives of the Hubble parameter are 
divergent, which implies divergence of some curvature invariants.
The big brake is a special example for a model from this class; it was
first considered in \cite{BGT} (see there the discussion after
Eq. (2.13)) and later discussed in detail in
\cite{we-tach}. It can arise in tachyonic
cosmological models \cite{sen} with a particular potential: at some 
finite moment of the cosmological evolution the universe stops at finite 
value of its cosmological radius with an infinite deceleration $\ddot{a}
\rightarrow -\infty$.
It was also noticed that the big-brake
singularity can arise in more simple cosmological models, 
such as a universe filled with a perfect fluid obeying the equation of
state $p = A/\rho$, where $A$ is a positive constant.
This equation of state was considered in \cite{anti-Chap} in the context of
wiggly strings (these are cosmic strings with small-scale wiggles
imposed on their dynamics). 
A fluid obeying this equation of state can be 
called ``anti-Chaplygin'' gas in analogy
with the gas with Chaplygin equation of state $p = -A/\rho$, which has
acquired some popularity in cosmology as candidate for unifying dark
energy and dark matter \cite{Chap,Chap1}. Independent of the possible
relevance of such a model for the real Universe, it has the merit of
showing that intriguing features can occur in the quantum
version, connected with the presence of a quantum phase at large
(instead of small) scale factor. Quantum effects at large cosmological
scales have previously been studied in the context of a classically
recollapsing quantum universe \cite{KieferZeh,Zeh}. 

Our paper is organized as follows:
In Sec. II we present a simple classical model exhibiting a big-brake
singularity. In Sec. III
the Wheeler--DeWitt equation for this model is studied and
approximate solutions describing wave packets are found. Their
behaviour demonstrates that the classical singularity is avoided. Sec. IV contains a discussion of the big-bang singularity.
Sec. V makes a comparison with the description of this model in loop
quantum cosmology. Sec. VI contains a discussion and an outlook.
Some technical details are relegated to an appendix.

\section{The classical big-brake model}

We consider a flat Friedmann--Lema\^{\i}tre universe filled with a
perfect fluid mimicked by a homogeneous scalar field. 
We require the fluid to obey
an ``anti-Chaplygin'' equation of state $p={A}/{\rho}$, where $p$ is
the fluid pressure and $\rho$ its energy density. In terms of the
scalar field, these read 

\be
p=\frac{\dot\phi^2}2-V(\phi)  , \quad \rho=\frac{\dot\phi^2}2+V(\phi) \ .
\ee
The corresponding action is

\bea
\lb{action}
S &=& \frac{3}{\kappa^2}\int\mathrm{d}t\ N\left(-\frac{a\dot{a}^2}{N^2}
+{\mathcal K}a-\frac{\Lambda a^3}{3}\right)\nonumber\\
&+&\frac{1}{2}\int\mathrm{d}t\ Na^3\left(\frac{\dot{\phi}^2}
{N^2}-2V(\phi)\right)\ ,
\eea
where $\kappa^2=8\pi G$, $N$ is the lapse function,
$\Lambda$ the cosmological constant, $V(\phi)$ a
potential of the field $\phi$, and
${\mathcal K}=0,\pm1$ is the curvature index; we set $c=1$.
Furthermore, we set $N=1$, so the  time parameter is the standard Friedmann cosmic time.
The action then becomes

\bea
\lb{action2}
S&=&\frac{3}{\kappa^2}\int{\mathrm d}t\ (-a\dot{a}^2+{\mathcal K}a-\frac{\Lambda}{3} a^3)\nonumber\\
&+&\frac12\int{\mathrm d}t\ (a^3\dot{\phi}^2-2a^3V(\phi))\ .
\eea
The canonical momenta are given by

\be
\lb{momenta}
\pi_a=-\frac{6a\dot{a}}{\kappa^2}\ , \quad \pi_{\phi}= a^3\dot{\phi}
\ .
\ee
The canonical Hamiltonian ${\mathcal H}$, which is constrained to
vanish, reads

\be
\lb{constraint}
{\mathcal H}=-\frac{\kappa^2}{12a}\pi_a^2+\frac{\pi_{\phi}^2}
{2a^3}+a^3\frac{\Lambda}{\kappa^2} + a^3V-\frac{3{\mathcal K}a}{\kappa^2}=0\ .
\ee
In the following, we restrict the analysis to flat cosmologies,
${\mathcal K}=0$, without cosmological constant, $\Lambda=0$.
The Hamiltonian constraint yields the Friedmann equation 
\be
\lb{Friedmann}
H^2=\frac{\kappa^2}{3}\rho=\frac{\kappa^2}{3}\left(\frac{\dot\phi^2}{2}+V(\phi)\right)\ .
\ee 
The fluid obeys a continuity equation,  
\be
\lb{KleinGordon1}
\dot\rho=-3H\left(\rho+p\right)\ ,
\ee
which in terms of the scalar field reads
\be
\lb{KleinGordon2}
\ddot\phi+3H\dot\phi+\frac{\d V}{\d\phi}=0 \ .
\ee
Using the equation of state, $p={A}/{\rho}$, \eqref{KleinGordon1}
can be easily solved for $\rho$ in terms of the scale factor,
\be
\lb{density}
\rho(a)=\sqrt{\frac{B}{a^6}-A}\ ,
\ee 
where $B>0$ is some integration constant, and we have chosen the
solution with $\rho\geq0$. Note that $\rho$ is well
defined only for $a<a_\star\equiv\left(B/A\right)^{1/6}$,
cf. Figure~1. As $a_\star$ is approached, the density goes to zero.
We note that $B$ has dimension mass squared, and $A$ has
dimension mass squared over length to the sixth power.

Using the result \eqref{density}, one gets from \eqref{Friedmann}:
\be
\lb{scalefactor}
\int_{a}^{a_\star} \frac{\d \tilde{a}}
{\left(\frac{B}{\tilde{a}^2}-A\tilde{a}^4\right)^{\frac14}}=
\frac{\kappa}{\sqrt{3}}\left(t_0-t\right)
\ ,
\ee
where $a(t_0)=a_\star$ (``big brake'') and $a(0)=0$ (``big bang'').
In order to calculate this integral, we substitute $z=(B/a^6-A)^{1/4}$,
with $0\leq z\leq \infty$. Then \eqref{scalefactor} becomes
\be
\lb{z-integral}
\int_0^z\d\tilde{z}\
\frac{\tilde{z}^2}{\tilde{z}^4+A}=\frac{\kappa\sqrt{3}}{2}(t_0-t)\ .
\ee
The integral on the left-hand side can be found in \cite{Gradshteyn1}.
For \eqref{z-integral} one then gets
\begin{eqnarray}
& & \frac{1}{4A^{1/4}\sqrt{2}}\left(\ln\frac{z^2-A^{1/4}z\sqrt{2}+A^{1/2}}
{z^2+A^{1/4}z\sqrt{2}+A^{1/2}}\right. \nonumber\\
& & \; \left. +2\arctan\frac{A^{1/4}z\sqrt{2}}{A^{1/2}-z^2}
+\pi\theta(z^2-A^{1/2})\right)\\&=&\frac{\kappa}{\sqrt{3}}(t_0-t)\ .
\end{eqnarray}
We have added the Heaviside $\theta$-function in order to make the
arctan-function continuous at the point $z^2=A^{1/2}$. 

For the total time that elapses from big bang to big brake one then gets
\begin{equation}
\lb{totaltime}
t_0=\frac{2}{\kappa\sqrt{3}}\int_0^{\infty}
\d z\ \frac{z^2}{z^4+A} = \frac{\pi}{\sqrt{6}\kappa A^{1/4}}
\end{equation}
The solution for $a(t)$ is shown in Figure~2.
A simple approximate solution can
be found in the vicinity of $a_\star$. To this end, we write
$a=a_\star-\Delta a$, which simplifies the above integral to
\be
\int_0^{\Delta a}\d\Delta a\frac1{a_\star(6\Delta
  a)^{\frac14}}=\sqrt{\frac{\kappa^2}{3}}\left(t-t_0\right)
\ ,
\ee 
yielding

\be
\Delta a(t)=\left[C(t_0-t)\right]^{\frac43}\ .
\ee
So we find for the scale factor and its derivatives

\be
a(t_0)=a_\star , \quad \dot a(t_0)=0, \quad \ddot a(t_0)=-\infty
\ .
\ee
At $t_0$, the evolution of the scale factor comes to a halt. Its `speed' is zero
due to an infinite negative acceleration. It is this peculiar feature
that gave the singularity its name, big-brake singularity.

The first and second time derivatives of the scale factor in terms of
the scale factor itself are given by 
simple expressions.
To this end, note that \eqref{scalefactor} can be differentiated with
respect to $a$, thus connecting $\dot a(t)$ with the scale factor
according to

\be
\lb{dota}
\frac{\d a}{\d
  t}=\sqrt{\frac{\kappa^2}{3}}a\left(\frac{B}{a^6}-A\right)^{\frac14}\ ,
\ee
cf. Figure~3.
Obviously, as $a\to a_{\star}$, $\dot a\to 0$. Differentiating again with
respect to time, one finds 

\be
\lb{ddota}
\frac{\d^2a}{\d
  t^2}=\frac{\kappa^2}{3}a\left(\frac{B}{a^6}-A\right)^{\frac12}\left[1-\frac{B}{4a^6}\left(\frac{B}{a^6}-A\right)^{-1}\right]\ ,
\ee
showing that $\ddot a(t)\to -\infty$ as $a\to a_{\star}$, cf. Figure~4.

What remains to be found, is an equation for $\phi$. As we
are interested in the quantum model, the solution in configuration
space, $\phi(a)$, suffices. This is obtained from

\be
\dot\phi^2=\rho+p\ ,
\ee 
using the equation of state and the Friedmann equation
\eqref{Friedmann}. The (exact) solution is

\be
\lb{classtrajectory}
\phi_{\mp}(a)=\mp\sqrt{\frac1{3\kappa^2}}\mathrm{artanh}
{\left(\sqrt{1-\frac{Aa^6}B}\right)}
\ ,
\ee
cf. Figure~5.
This is only consistent if the potential is chosen to be

\be
V(\phi)=V_0\left(\sinh{\left(\sqrt{3\kappa^2}|\phi|\right)}-\frac1{\sinh{\left(\sqrt{3\kappa^2}|\phi|\right)}}\right)
\ .
\ee
Given the trajectories $\phi(a)$ and $a(t)$, the latter in explicit
form only in the
vicinity of the singularity, the classical model is thus fully
described. Note that $V_0=\sqrt{{A}/{4}}$. From \eqref{totaltime} we
find for the total lifetime of this model universe the expression
\be
t_0\approx 7\times 10^{2}\frac{1}{\sqrt{V_0\left[\frac{\rm g}{{\rm
          cm}^3}\right]}}\ {\rm s}\ .
\ee
This lifetime is much bigger than the current age of our Universe if
\begin{displaymath}
V_0\ll 2.6\times 10^{-30}\ \frac{\rm g}{{\rm cm}^3}\ ,
\end{displaymath}
which is, of course, a reasonable result because the critical value of
$V_0$ just corresponds to the scale of the observed dark-energy density.

\begin{figure}
\begin{minipage}[t]{0.45\linewidth}
\scalebox{1.0}{\hspace{-20mm}\includegraphics[angle=0]{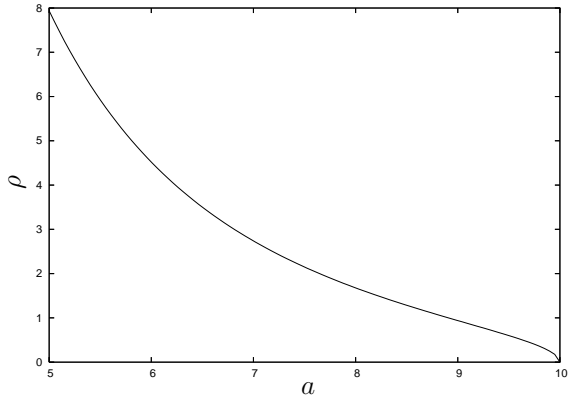}}
\caption{\lb{rho} Evolution of the energy density $\rho$ of the scalar field
  with scale factor $a$.}
\end{minipage}
\end{figure}
\begin{figure}
\begin{minipage}[t]{0.45\linewidth}
\scalebox{1}{\hspace{-20mm}\includegraphics[angle=0]{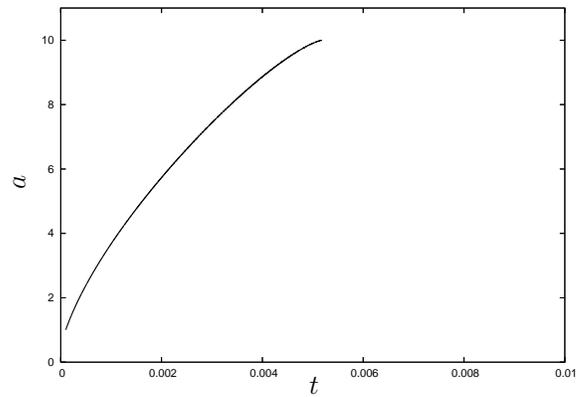}}
\caption{\lb{a} Evolution of the scale factor over cosmic Friedmann
  time $t$.}
\end{minipage}
\end{figure}
\begin{figure}
\begin{minipage}[t]{0.45\linewidth}
\scalebox{1}{\hspace{-20mm}\includegraphics[angle=0]{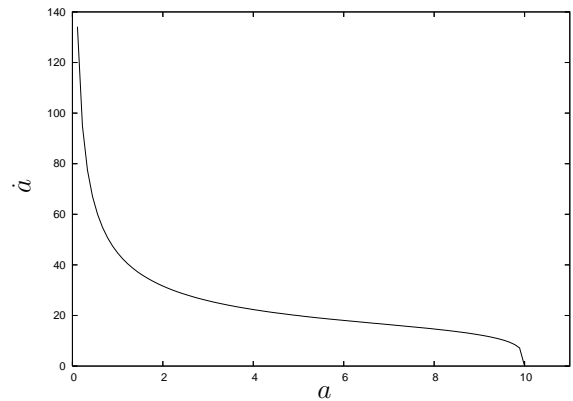}}
\caption{\lb{da} Dependence of the derivative of the scale factor on
  the scale factor itself.}
\end{minipage}
\end{figure}
\begin{figure}
\begin{minipage}[t]{0.45\linewidth}
\scalebox{1}{\hspace{-20mm}\includegraphics[angle=0]{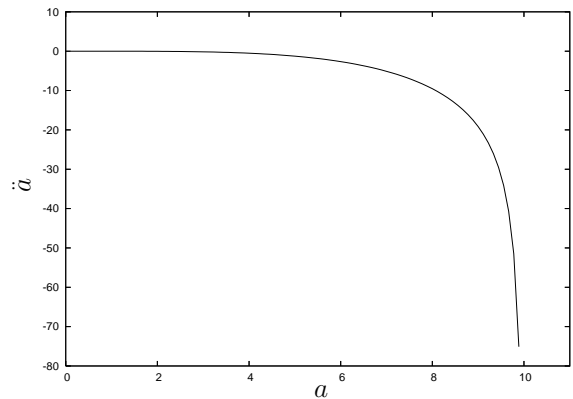}}
\caption{\lb{dda} Cosmic acceleration depicted over $a$.}
\end{minipage}
\end{figure}
\begin{figure}
\begin{minipage}[t]{0.45\linewidth}
\scalebox{1}{\hspace{-20mm}\includegraphics[angle=0]{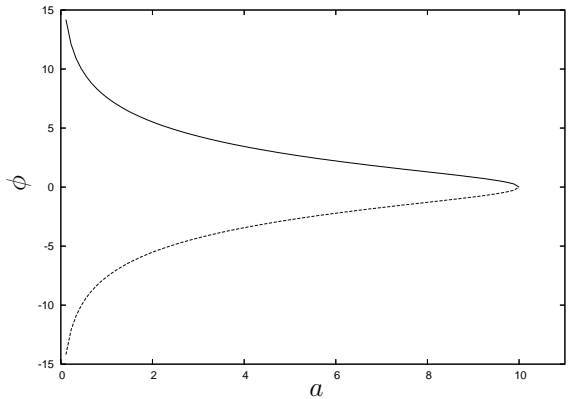}}
\caption{\lb{phi} Classical trajectory in configuration space.}
\end{minipage}

\end{figure}
\section{The quantum big-brake model}
\label{Quantization}
\subsection{Wheeler--DeWitt equation}

Quantization is carried out in the canonical approach. Implementing
the Hamiltonian constraint via Dirac's constraint quantization, one
arrives at the Wheeler--DeWitt equation\\

\bea
\lb{WheelerDeWitt1}
& &\frac{\hbar^2}{2}\left(\frac{\kappa^2}{6}\frac{\del^2}{\del\alpha^2}-\frac{\del^2}{\del\phi^2}\right)\Psi\left(\alpha,
  \phi\right)\nonumber\\
&+&V_0e^{6\alpha}\left(\sinh{\left(\sqrt{3\kappa^2}|\phi|\right)}-\frac1{\sinh{\left(\sqrt{3\kappa^2}|\phi|\right)}}\right)\Psi\left(\alpha,
  \phi\right)\nonumber\\ &=&0,
\eea

where $\alpha\equiv\ln{a}$ and the Laplace--Beltrami factor ordering
has been employed. 
As we are interested in the behaviour in the vicinity of the big-brake
singularity, where $\phi$ is small, it is sufficient to approximate
the potential there.
We find 

\be
\lb{WheelerDeWitt2}
\frac{\hbar^2}2\left(\frac{\kappa^2}{6}\frac{\del^2}{\del\alpha^2}-\frac{\del^2}{\del\phi^2}\right)\Psi\left(\alpha,
  \phi\right)-\frac{\tilde{V_0}}{|\phi|}e^{6\alpha}\Psi\left(\alpha,
  \phi\right)=0
\ ,
\ee
where $\tilde{V_0}={V_0}/{3\kappa^2}$.
\subsection{Born--Oppenheimer approximation to the Wheeler--DeWitt equation}

Equation \eqref{WheelerDeWitt2}
 can be solved, at least approximately, making the ansatz $\Psi\left(\alpha,
  \phi\right)=\sum_k C_k(\alpha)\varphi_k(\alpha, \phi)$, where
$\varphi_k(\alpha, \phi)$ is the solution of

\be
\lb{Coulombeqn}
-\left(\frac{\hbar^2}2\frac{\del^2}{\del\phi^2}+\frac{\tilde{V_0}}{|\phi|}e^{6\alpha}\right)\varphi_k(\alpha,
\phi)=E_k(\alpha)\varphi_k(\alpha, \phi)
\ ,
\ee
cf. also \cite{packet1}, where a similar ansatz was made.
We recognize that this is the radial part of the
time-independent Schr\"odinger equation
for a particle in a Coulomb potential with $l=0$ and the wave function
$r\varphi_k$.  
Thus, the normalizable solutions are given by
\be  
\lb{phi_k}
\varphi_k(x_k)=N_kx_ke^{-\frac{x_k}{2}}\mathrm{L}^1_{k-1}(x_k)\ ,
\ee
where $x_k=2\sqrt{-\frac{2E_k(\alpha)}{\hbar^2}}|\phi|$, and
$\mathrm{L}^1_{k-1}(x_k)$ denote the associated Laguerre
polynomials; $N_k=1/k^{\frac32}$ is the normalization factor;
$k\in{\mathbb N}$ 

The choice of the normalizable solution to \eqref{Coulombeqn} is
  enforced through the condition on the wave function imposed for
  large $\vert\phi\vert$, cf. Sec. \ref{BigBangSingularityAvoidance}.
 The exact normalizable solution to \eqref{Coulombeqn} with the exact
 potential possesses a discrete spectrum; coincidence with the
 behaviour at small $\vert\phi\vert$ is thus only achieved if the
 normalizable solution \eqref{phi_k} is selected because the
 non-normalizable solutions have a continuous spectrum.

Note that $\varphi_k(x_k)\to 0$ for $\vert\phi\vert \to 0$, since 
$\mathrm{L}^1_{k-1}(0)=k$. 
To simplify notation, introduce
$Z(\alpha)\equiv\hbar^2/V_{\alpha}$ and
$V_\alpha\equiv\tilde{V_0}e^{6\alpha}$. Then, $x_k=2|\phi|/Z(\alpha)k$.
The functions $\varphi_k(x_k)$ are orthogonal such
that\footnote{The validity of this relation is
clear from the property of the $\varphi_n$ being eigenfunctions of a
Hermitian operator; its direct verification
is discussed in \cite{Dunkl}.}
\be
\lb{ortho}
\int\d\phi\ \varphi_k(x_k)\varphi_l(x_l)=Z(\alpha)\delta_{kl}\ .
\ee
The energy eigenvalues are 

\be
\lb{energyeigenvalues}
E_k(\alpha)=-\frac{V_\alpha^2}{2\hbar^2k^2}
\ .
\ee
Inserting this ansatz in \eqref{WheelerDeWitt2} and carrying out a
Born--Oppenheimer approximation,
the resulting equation for $C_k(\alpha)$ becomes

\be
\lb{Besseleqn}
\ddot
C_k(\alpha)-\frac{6{V_\alpha}^2}{\hbar^4k^2\kappa^2}C_k(\alpha)=0\ ,
\ee
where dots denote derivatives with respect to $\alpha$. Thus $C_k$ is
given by

\be
\lb{C_k}
C_k(\alpha)=c_1\mathrm{I}_0\left(\frac{1}{\sqrt{6}}\frac{V_{\alpha}}{\hbar^2k\kappa}\right)+c_2\mathrm{K}_0\left(\frac{1}{\sqrt{6}}\frac{V_{\alpha}}{\hbar^2k\kappa}\right)\ ,
\ee
where $\mathrm{I}_0$, ${\mathrm K}_0$ denote modified Bessel functions of first and
second kind, respectively. As a boundary condition, we require that
the solution should vanish in the classically forbidden region,
$a>a_\star$. Therefore, $c_1=0$ and only the MacDonald function ${\mathrm K}_0$
remains as solution.
On the level of the Born--Oppenheimer approximation, the complete
solution is therefore given by

\bea
\lb{BOsolution}
\Psi\left(\alpha,\phi\right)=
\sum_{k=1}^{\infty}A(k)N_k\mathrm{K}_0
\left(\frac{1}{\sqrt{6}}\frac{V_{\alpha}}{\hbar^2k\kappa}\right)\nonumber\\
\times\left(2\frac{V_\alpha}{k}|\phi|\right)
e^{-\frac{V_\alpha}{k|\phi|}}\mathrm{L}^1_{k-1}
\left(2\frac{V_\alpha}{k}|\phi|\right)\ .
\eea

\subsection{Derivation of classical equations of motion from the
  principle of constructive interference}

To derive a phase from this expression, approximate \eqref{Coulombeqn}
and \eqref{Besseleqn} further by a WKB-approximation. Making the
ansatz $\varphi_k(\alpha, \phi)=e^{\frac i\hbar S_{k0}^{\phi}(\alpha, \phi)}$
in \eqref{Coulombeqn}, $C_k(\alpha)=e^{\frac i\hbar
  S_{k0}^{\alpha}(\alpha)}$ in \eqref{Besseleqn}, one obtains to
zeroth order in $\hbar$ the Hamilton--Jacobi equation for the $\phi$- and $\alpha$-part, respectively. Integration yields for $S_{k0}^{\phi}(\alpha, \phi)$:

\bea
\lb{S_phi}
S_{k0}^{\phi}(\alpha, \phi)
=\hbar
k\left[\mathrm{arcsin}\left(1-\frac{V_{\alpha}|\phi|}{\hbar^2k^2}\right)
-\frac{\pi}{2}\right]\nonumber\\-
\sqrt{2V_{\alpha}|\phi|}
\sqrt{1-\frac{V_{\alpha}|\phi|}{2\hbar^2k^2}}-\frac{\pi}{4}\ ,
\eea
in which the Langer boundary
condition at the $\alpha$-dependent turning point
$\phi_t\left(\alpha\right)=2\hbar^2k^2/V_{\alpha}$ has been employed.
From \eqref{Besseleqn}, no phase results. This coincides with
the limit $\hbar\to\ 0$ in \eqref{C_k}, as
$\lim_{x\to\infty}\mathrm{K}_0\left(x\right)\approx\sqrt{\frac{\pi}{2x}}e^{-x}$.
So $S_{k0}^{\phi}(\alpha,
\phi)$ constitutes the entire phase. 

The classical equations of motion should
follow from the phase through the principle of constructive
interference, $\frac{\del S_{k0}^{\phi}}{\del k}|_{k=\bar k}=0$:

\bea
\lb{constructiveinterference}
\frac{\del S_{k0}^{\phi}}{\del k}|_{k=\bar k}&=&
\hbar\left[\mathrm{arcsin}\left(1-\frac{V_{\alpha}|\phi|}{\hbar^2k^2}\right)
-\frac{\pi}{2}\right]\nonumber\\
&+&\frac{\sqrt{2V_{\alpha}|\phi|}}{k}
\sqrt{1-\frac{V_{\alpha}|\phi|}{2\hbar^2k^2}}\nonumber\\
&\stackrel{!}{=}&0\ ,
\eea
Here, $\bar
k=\sqrt{\frac{\tilde{V_0}}{\sqrt{3\kappa^2}}}\frac{a_{\star}^3}{\hbar}$.
This constant arises under the conditions that, firstly, $k$ and
  so also $\bar k$ have to be dimensionless, and that, secondly, the
  only constants of the model are $V_0$ (or $\tilde{V_0}$),
  $a_{\star}$ (or $A$ and $B$), $\hbar$ and $\kappa$.
With this choice, \eqref{constructiveinterference} simplifies to

\bea
\frac{\del S_{k0}^{\phi}}{\del k}|_{k=\bar k}=
\hbar\left[-\mathrm{arccos}\left(1-\left(\frac{a}{a_{\star}}\right)^6|\phi|\right)\right.\nonumber\\
   \ \left.
+\left(\frac{a}{a_{\star}}\right)^3\sqrt{2|\phi|-\left(\frac{a}{a_{\star}}\right)^6\phi^2}\right]\ .
\eea
For the classical trajectory, \eqref{classtrajectory}, this is

\bea
\lb{construcitveinterference2}
\frac{\del S_{k0}^{\phi}}{\del k}|_{k=\bar k}=
\hbar\left[-\mathrm{arccos}\left(1-\frac{|\phi|}{\mathrm{cosh}^2
\left(\sqrt{3\kappa^2}|\phi|\right)}\right)\right.\nonumber\\  \ \left.
+\frac{\sqrt{|\phi|}}{\mathrm{cosh}\left(\sqrt{3\kappa^2}|\phi|\right)}
\sqrt{2-\frac{|\phi|}{\mathrm{cosh}^2\left(\sqrt{3\kappa^2}|\phi|\right)}}
\right]\ .
\eea
But the classical equation of motion was derived using the full
potential. The quantum theory uses an approximation to the original
potential which is valid up to order
$\mathcal{O}\left(|\phi|^{\frac32}\right)$ for small $\phi$. Applying
the same approximation to \eqref{construcitveinterference2}, one finds

\be
\frac{\del S_{k0}^{\phi}}{\del k}|_{k=\bar
  k}=\hbar\hspace{1mm}\mathcal{O}\hspace{-1.5mm}
\left(|\phi|^{\frac32}\right)\ ,
\ee 
and so the classical solution \eqref{classtrajectory} satisfies the
condition for constructive interference with the above choice for
$\bar k$ for small $\phi$, which is consistent with the approximation
of the potential in \eqref{WheelerDeWitt2}.

There is, of course, also the question whether the Born--Oppenheimer
approximation employed in the last subsection is a feasible
approximation. 
We show in Appendix~A that this approximation is fulfilled in the limit $a\to
a_{\star}$, which is just the region under consideration here.


\subsection{Singularity avoidance}

Wave packets in quantum cosmology have been constructed in order to
study aspects of the quantum-to-classical correspondence, in
particular the validity of the semi-classical approximation
\cite{OUP,packet1,packet2}. They are also useful in order to provide a
consistent picture of the pre-big-bang to post-big-bang transition in
quantum string cosmology \cite{DK}.
 such a construction is also useful in the
study of singularity avoidance.

Wave packets constructed from the solutions of \eqref{WheelerDeWitt2}
are of the general form
\be
\Psi(\alpha, \phi)=\sum_{k=1}^{\infty}A_kC_k(\alpha)\varphi_k(\alpha, \phi)\ .
\ee
We can choose initial conditions on a hypersurface
$\alpha=\alpha_0$. Here, it suffices to fix the values $\Psi(\alpha_0,
\phi)$ and $\frac{\del\Psi(\alpha,
  \phi)}{\del\alpha}|_{\alpha=\alpha_0}$. As 
for the chosen normalizable solution \eqref{phi_k} $\varphi_k(\alpha, \phi)$
vanishes at $\phi=0$ for all $k$ and $\alpha$, the wave packet is zero
there. This is, of course, independent of the initial conditions.
But the classical singularity occurs at $\phi=0$. So out of these
solutions, no wave packet can be constructed which does {\em not}
vanish at the classical singularity. 
Taking $\alpha$ as an internal time variable, one can calculate the probability
distribution, 

\be
|\Psi|^2(\alpha_0, \phi)=\sum_{l,k}A_k A_l
C_k(\alpha_0)C_l(\alpha_0)\varphi_l(\alpha_0, \phi)\varphi_k(\alpha_0,
\phi)\ ,
\ee
for each `instant of time' $\alpha_0$. It is obvious that 
$|\Psi|^2(\alpha_0, 0)=0$ at $\phi=0$. We emphasize that this is a
consequence of the choice of \eqref{phi_k}.

To manifest the elimination of the classical singularity on the
quantum level, also expectation values have been employed, see, for
example, \cite{Ashtekar}.
Before calculating the expectation value for $\vert\phi\vert$ for this model using the inner product \eqref{ortho}, recall that the avoidance of the singularity of the Coulomb potential in ordinary quantum mechanics is caused by a lowest bound on the energy due to quantization. This again leads to a minimal radius for the `trajectory' of the electron.

Analogously to the Coulomb potential in ordinary quantum mechanics,
the energy (of the matter component) in our model is also bounded from
below. The minimal energy, given by \eqref{energyeigenvalues} for
$k=1$, corresponds to a minimal `radius', that is, 
to a minimal value for $\vert\phi\vert$. This is given by

\ben
\lb{expectationvalue}
\langle|\phi_k|\rangle(\alpha)&=&\left[C_k(\alpha)\right]^2
\frac32\left[Z(\alpha)\right]^2k^2\\
&=&\left[K_0\left(\frac{1}{\sqrt6}\frac{V_{\alpha}}{\hbar^2k\kappa}\right)\right]^2
\frac{3\hbar^4}{2V_{\alpha}^2}k^2\ ,
\een
for $k=1$. The classical singularity lies at $\alpha=\alpha_{\star}$. In this case the minimal energy is given by

\be
E_1(\alpha_{\star})=-\frac{V_{\alpha_{\star}}^2}{2\hbar^2}\ ,
\ee
and the expectation value for $\vert\phi\vert$ is consequently given by $\langle|\phi_1|\rangle(\alpha_{\star})$. The boundedness of the energy here prevents the scalar field to evolve to the singularity, $\vert\phi\vert=0$, in this case as well.

Note that for $\alpha\to\infty$, the energy is no longer bounded. In this case $\langle|\phi_1|\rangle\to 0$, cf. \eqref{expectationvalue}. 
Of course, one should keep in mind that the expectation value in quantum cosmology has no interpretation in terms of measurement results as it has in conventional quantum theory. 

\subsection{Construction of wave packets}

Apart from the avoidance of the singularity, we want to study
semi-classical and quantum regimes of the model. To this end, we
construct semi-classical wave packets and study their
behaviour. Especially we are interested in the regions of
configuration space where these packets spread (if they spread at
all). 

We want $\Psi(\alpha_0, \phi)$ to be a Gaussian centered at
$\phi_0$ with width
$\sqrt{\frac{Z_0}{2}}$, where $Z_0\equiv
Z(\alpha_0)$. The center $\phi_0$ should be the value of the classical 
trajectory at $\alpha_0$. Note that we have two classical solutions, 
$\phi_+$ and $\phi_-$, see \eqref{classtrajectory}.\footnote{The case
  with two Gaussians is the most general case. One may, of course,
  wish to choose only one Gaussian in order to represent only one
  branch of the classical solutions by a wave packet.} 
So in fact, we have to construct two Gaussians, 
one centered at $\phi_0$, the other at
$-\phi_0$ and superpose both. Write therefore

\be
\Psi(\alpha_0, \phi)=\Psi_-(\alpha_0, \phi)+c_1\Psi_+(\alpha_0, \phi)\ ,
\ee
where $\Psi_+$ denotes the part of the wave packet being centered around
$\phi_0$ and $\Psi_-$ the part centered around $-\phi_0$ at initial
`time' $\alpha_0$.
\\
The calculation of the wave packet will employ only the WKB solution of
\eqref{Besseleqn}. With suitable initial conditions, it reads

\be
\lb{WKB}
C_k(\alpha)=\left(\frac{e^{6\alpha_0}}{e^{6\alpha}}\right)^{\frac12}
\exp{\left[-\frac16\frac{\tilde{V_0}}{\sqrt{2\hbar^2
        k^2}}\sqrt{\frac{6}{\kappa^2}}\left(e^{6\alpha}-e^{6\alpha_0}\right)\right]}\ .
\ee
Introducing $\tau\equiv e^{6\alpha}$ (and denoting $\tau_0\equiv
e^{6\alpha_0}$),

\be
C_k(\tau)=\left(\frac{\tau_0}{\tau}\right)^{\frac12}
\exp{
\left[-\frac16\frac{\tilde{V_0}}{\sqrt{2\hbar^2 k^2}}\sqrt{\frac{6}{\kappa^2}}\left(\tau-\tau_0\right)\right]}\ .
\ee
Start with the $\Psi_+$--part of the wave packet.
We here find the requirement

\be
\Psi_+(\alpha_0, \phi)=\sum_{k=1}^{\infty}A_k^+\varphi_k(\alpha_0,
\phi)
\stackrel{!}{=}e^{-\frac{(\phi-\phi_0)^2}{Z_0}}\ .
\ee
Decomposing the Gaussian into the $\varphi_k(\alpha_0, \phi)$, one
obtains for the coefficients the somewhat lengthy expression

\bea
A_k^+=&\frac{N_k}{k}
\exp{\left[-\frac{\phi_0^2}{Z_0}+\frac1{2Z_0}\left(\frac1{2k}-\phi_0\right)^2\right]}\times\nonumber\\
&\sum_{m=0}^{k-1}(-1)^m(m+1)\frac{(k!)^2}{(k-m-1)!(m+1)!}\nonumber\\
&\left(\sqrt{\frac2Z_0}\frac1k\right)^m
\mathrm{D}_{-(m+2)}
\left[\sqrt{\frac2Z_0}\left(\frac1{2k}-\phi_0\right)\right]\ ,
\eea
where $D_m(x)$ denote parabolic cylinder functions.
Note that this expansion in $\varphi_k$ cannot be performed at
$\phi=0$. Here, $\varphi_k(\alpha, \phi=0)=0$ for all $k$ as remarked
above.

The amplitude of $\Psi_-$ is obtained in a similar way (or
by just substituting $-\phi_0$ for $\phi_0$). The
solution is

\bea
A_k^-=&\frac{N_k}{k}
\exp{\left[-\frac{\phi_0^2}{Z_0}+\frac1{2Z_0}\left(\frac1{2k}+\phi_0\right)^2\right]}\times\nonumber\\
&\sum_{m=0}^{k-1}(-1)^m(m+1)\frac{(k!)^2}{(k-m-1)!(m+1)!}\nonumber\\
&\left(\sqrt{\frac2Z_0}\frac1k\right)^m
\mathrm{D}_{-(m+2)}
\left[\sqrt{\frac2Z_0}\left(\frac1{2k}+\phi_0\right)\right]\ .
\eea
So the wave packet is given by

\be
\Psi(\alpha, \phi)=\sum_{k=1}^{\infty}
\left[A_k^++c_1A_k^-\right]C_k(\alpha)\varphi_k(\alpha, \phi)\ .
\ee
The total probability for the wave packet is calculated via

\bea
\int\d\phi\hspace{1mm}\vert\Psi\vert^2
=&\frac{\tau_0\hbar^2}{\tilde{V_0}}\frac1{\tau^2}
\sum_{k=1}^{\infty}\left[A_k^++c_1A_k^-\right]^2\nonumber\\
&\exp{\left(-\frac13\frac{\tilde{V_0}}{\sqrt{2\hbar^2k^2}}\sqrt{\frac{6}{\kappa^2}}(\tau-\tau_0)\right)}\ .
\eea
Probability is thus not conserved with respect to internal `time'
$\tau$, as expected \cite{OUP}. 
Choose the normalization of the wave packet such that at
$\alpha_0$, $\int\d\phi\hspace{1mm}\vert\Psi\vert^2=1$. Then,

\be
\lb{Wavepacketplot}
\Psi(\alpha, \phi)=\frac1C\sum_{k=1}^{\infty}
\left[A_k^++c_1A_k^-\right]C_k(\alpha)\varphi_k(\alpha, \phi)\ ,
\ee
where the normalization factor is given by

\be
C\equiv\sqrt{\frac{\hbar^2}{\tilde{V_0}\tau_0}
\sum_{k=1}^{\infty}\left[A_k^++c_1A_k^-\right]^2}\ .
\ee
A plot of the wave packet is shown in Figure~6. We recognize that 
the wave function is peaked around the two branches of the 
classical trajectory in
configuration space, but goes to zero if the region of the classical
big-brake singularity, $a\to a_{\star}$, is approached. In this sense
the classical singularity is avoided in the quantum theory.
This is a consequence of the choice of the normalizable solution (\ref{phi_k}), which vanishes at $\phi=0$ (the region of the big-brake singularity).
Moreover, we find that the wave packet does not spread along the
classical trajectory. 

\begin{figure}
\scalebox{0.6}{\hspace{-7mm}\includegraphics[angle=0]{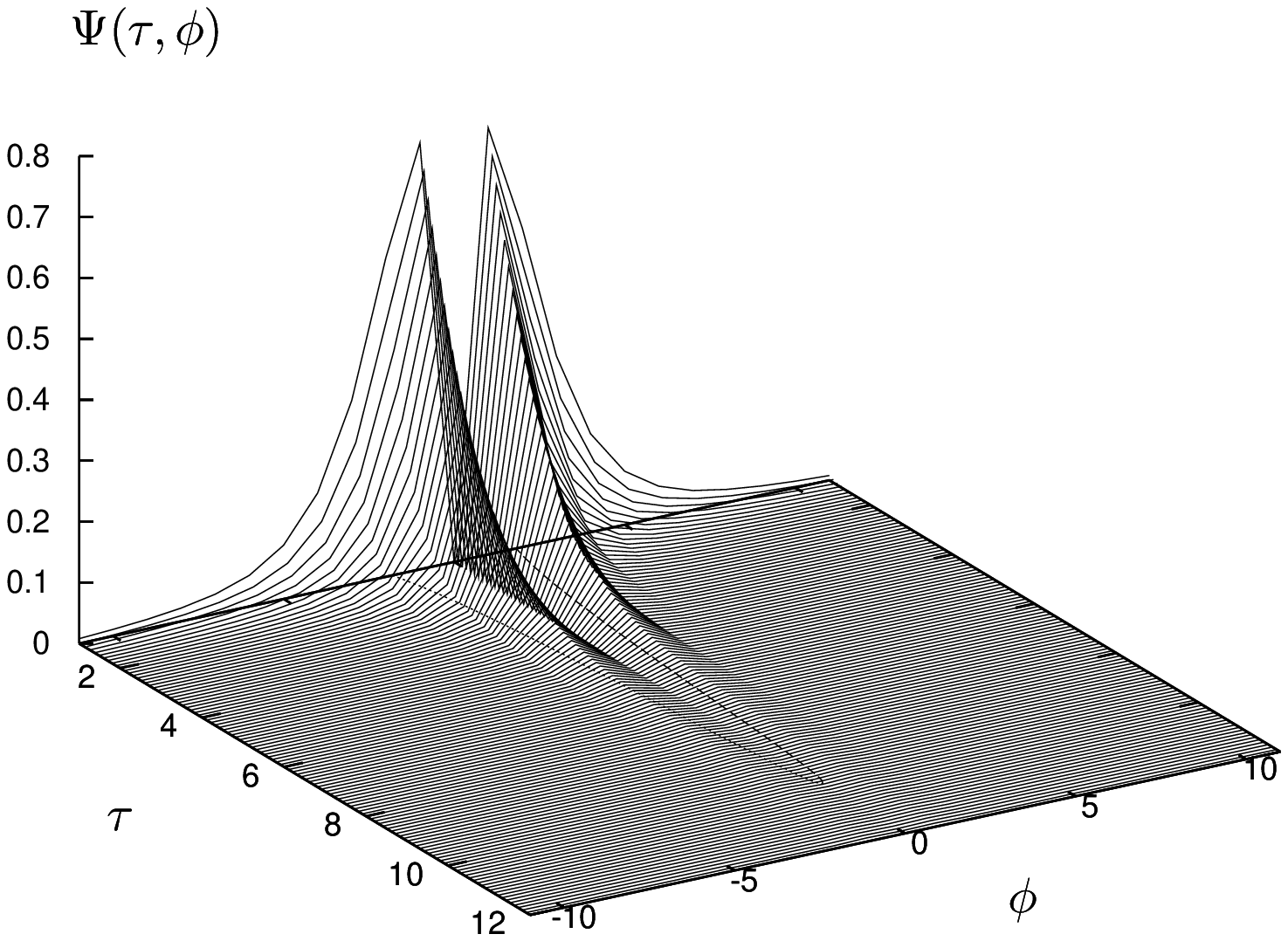}}
\caption{This plot shows the wave packet. 
It follows classical trajectories with initial values $a_0=1$ and
  $\phi_0\approx 0.88$. The classical trajectories are
depicted in the $(\tau,\phi)$-plane; recall $\tau=a^6$.\\
This corresponds to a singularity occuring at
$a_{\star}=10^{\frac16}$. Parameter values are $\tilde{V_0}=1$, $\hbar=1$ and $c_1=1$, cf. \eqref{Wavepacketplot}. 
Summation was carried out up to $k=50$.}
\end{figure}

\section{Remarks on Big-Bang singularity}
\subsection{Solution to the Wheeler--DeWitt equation}
So far, only the big-brake singularity of the model was
considered. But the model possesses a  second singularity. Namely, its
evolution starts with a big bang: as
$a\to 0$, one has $\vert\phi\vert\to\infty$. 
Thus one can approximate the potential by
an exponential in the vicinity of this singularity. Choosing units such that $\kappa^2=6$, one obtains the
following form of the Wheeler--DeWitt equation:

\be
\frac{\hbar^2}{2}\left(\frac{\del^2}{\del\alpha^2}-\frac{\del^2}{\del\phi^2}\right)\Psi+\frac{\tilde{V_0}}{2}e^{6\alpha+3\sqrt{2}|\phi|}\Psi=0\ .
\ee
Introducing coordinates $z_1=\alpha+|\phi|$, $z_2=\alpha-|\phi|$, the
Wheeler--DeWitt equation becomes

\be
\hbar^2\frac{\del^2}{\del z_1\del z_2}\Psi=f(z_1,z_2)\Psi\ .
\ee
One can now find coordinates so that the function on the right-hand
side cancels. One is left with

\be
\lb{50}
\hbar^2\left(\frac{\del^2}{\del u^2}-\frac{\del^2}{\del
    v^2}\right)\Psi
+\Psi=0\ ,
\ee
where

\bea
u(\alpha,\phi)=\frac{2\sqrt{\tilde{V_0}}}{3}e^{3\left(\alpha+\frac{1}{\sqrt2}|\phi|\right)}\left[\cosh
X-\frac{1}{\sqrt2}\sinh X\right]\hspace{-1.5mm}\ ,\\
v(\alpha,\phi)=\frac{2\sqrt{\tilde{V_0}}}{3}e^{3\left(\alpha+\frac{1}{\sqrt2}|\phi|\right)}\left[\sinh
  X-\frac{1}{\sqrt2}\cosh X\right]\hspace{-1.5mm}\ ,
\eea
and $X\equiv3\left(|\phi|+\frac{1}{\sqrt2}\alpha\right)$.
A solution to this equation can be found from the WKB--ansatz
$\Psi=\int \d k A(k) e^{\pm\frac{i}{\hbar}S_{0k}}$. 
Inserting this ansatz into (\ref{50})
yields the Hamilton--Jacobi equation of which an exact
solution is given by $S_{0k}=ku-\sqrt{k^2-1}v$. Of course,
the Hamilton--Jacobi equation is also solved by actions with different
signs in front of $u$ and $v$. These are obtained from the one chosen
above through rotations in the $(u,v)$-plane. As $u>0$, only two
solutions can be mapped onto each other. 

\subsection{Recovery of classical trajectories}
The classical
trajectory in the vicinity of the big bang is recovered using the
principle of constructive interference 
$\frac{\d S_{0k}}{\d k}|_{k=\bar{k}}=0$. For $\bar{k}=\sqrt2$ one finds
$\phi(\alpha)=\pm\frac{1}{\sqrt2}\alpha$. This is just the classical
trajectory obtained from \eqref{classtrajectory} in the limit
$|\phi|\gg 1$ with initial condition $B=\frac A4$ and fixed $A$.

\subsection{Construction of wave packets}
We get the
following exact wave-packet solution to the Wheeler--DeWitt equation:
\be
\Psi(u,v)=\int dk\,
A(k)\left(C_1e^{\frac{i}{\hbar}(ku-\sqrt{(k^2-1)}v)}+C_2\hspace{2mm} c.c.\right)\ ,
\ee
where $c.c.$ denotes the complex conjugate of the precee-ding term.
By construction, the classical trajectories can be recovered from this
equation through the principle of constructive interference. Choosing as
amplitude a Gaussian with width $\sigma$ centered around $\bar{k}$, 

\bdm
A(k)=\frac{1}{(\sqrt{\pi}\sigma\hbar)^{1/2}}e^{-\frac{(k-\bar{k})^2}
{2\sigma^2\hbar^2}}\ ,
\edm
and taking $C_1=C_2$ for definiteness, one obtains wave packets of the form

\bea
\label{wavepacket}
\psi(u_\ell,v_\ell)\approx C_1\pi^{1/4}
\sqrt{\frac{2\sigma\hbar}{1-i\sigma^2\hbar S_0^{\prime\prime}}}\nonumber\\ 
\exp\left(\frac{iS_0}{\hbar}-\frac{S_0^{\prime 2}}{2(\sigma^{-2}
-i\hbar S_0^{\prime\prime})}\right)
+ \mathrm{c.c.}\ ,
\eea
where a Taylor expansion of $S_{0k}$ has been carried out around
$\bar{k}$ (primes denoting derivatives with respect to $k$) and the terms of the order $(k-\bar{k})^3$ in  the exponent have been
neglected. (For simplicity, in this expression
$S_{0k}(\bar{k})\equiv S_0$.) This can be done if the Gaussian is strongly peaked around
$\bar{k}$, that is, if $\sigma$ is sufficiently small.
Since $S_{0k}'(\bar{k})=0$ gives the classical trajectory, the packet
is peaked around it. 

\subsection{Singularity avoidance 
\lb{BigBangSingularityAvoidance}}

Due to the fact that $u>0$, two inequivalent actions exist. Apart from the wave
packet constructed from $S_{0k}=ku-\sqrt{k^2-1}v$, one gets a
second wave packet constructed from
$S_{0k}=-ku-\sqrt{k^2-1}v$. Moreover, the entire $(\alpha,\phi)$
plane is mapped into only a quarter of the $(u,v)$ plane. One
would therefore require the wave packet to vanish on the
boundary of the physical region. The only solution satisfying this
requirement is naturally the trivial one.
To get a non-trivial solution, one has to lessen the boundary
condition and require $\Psi=0$ only at the origin of the $(u,v)$ plane.
The fact that the wave packet does not vanish at the $u=0$ and
$v=0$ line is due to the non-normalizability of the wave packet in
both $\alpha$ and $\phi$, which in turn has its origin in the fact that
the approximation to the classical trajectory for large $|\phi|$ has no turning point.\\ 
The implementation of the condition of normalizability results in a wave
packet which vanishes at the big-bang singularity, $\Psi\to 0$ as
$\alpha\to -\infty$, and spreads for large $\alpha$.
This is equivalent to the condition $\Psi\to 0$ as
$\vert\phi\vert\to\infty$. The condition implied for large
$\vert\phi\vert$ in the vicinity of the big bang thus implies and
justifies the normalization condition imposed in the derivation of the
solution to the Wheeler--DeWitt equation in the vicinity of the big
brake, cf. \eqref{Coulombeqn}. We thus impose basically two conditions
on the wave function. The first one is that $\Psi\to 0$ when
$\vert\phi\vert\to\infty$, resulting in a normalization condition 
for the approximate solution in
the vicinity of the big-brake singularity and the elimination of the
big-bang singularity. The second condition is to require $\Psi\to 0$ as
$a\to\infty$ to ensure the existence of wave packets that follow the
classical trajectory. Upon matching the wave function in the two
regimes, one would expect quantization conditions as observed e.g. in
\cite{packet1} or \cite{packet2}.
The big-bang singularity does therefore not exist in the quantum
theory.\\
The method employed in this section mirrors the calculation carried
out in \cite{MCB}. 
The picture one obtains is thus the following. For large $|\phi|$, the
wave packet vanishes and so does the wave packet for small
$|\phi|$. In the intermediate region, the packet is peaked around the
corresponding approximation to the classical trajectory, cf. Figure~7.

\begin{figure}
\scalebox{0.7}{\hspace{-5mm}\includegraphics[angle=0]{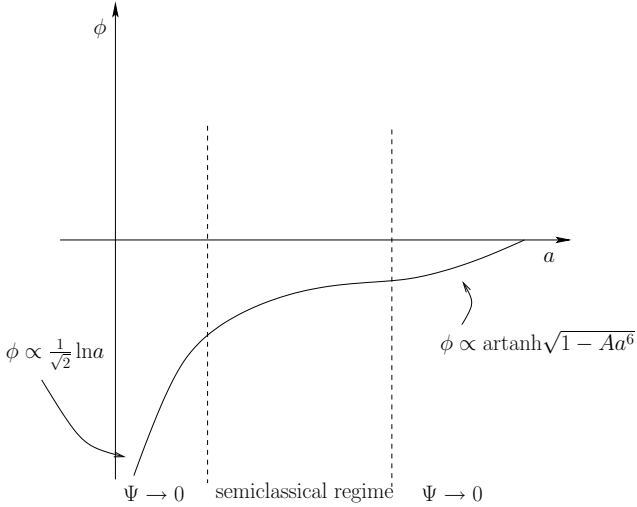}}
\caption{\lb{matching} The different regions of the wave packets and
  the classical trajectories are shown.}
\end{figure}

\section{Relation to Loop Quantum Cosmology}

As discussed in the introduction, there is a whole class of difference
operators 
in the current loop quantum cosmology literature. 
The ambiguity stems not only from the freedom to choose a factor
ordering, but also from the fact that the Hamiltonian 
constraint contains a curvature term which, 
when expressed in terms of holonomies,
is given by a limiting procedure. This limiting procedure consists of
shrinking an area to zero. But in loop quantum gravity there is
a smallest area and thus the limit is heuristically reduced to
\be
\mathrm{Area}\to\Delta\equiv\text{minimal area} \ .
\ee
 There are at least two ways to implement this. First, one can
 send the side length of the area to the value
 $\mu\to\mu_0\equiv\Delta^{\frac12}$ \cite{Bojowald}. Then $\mu_0$ is just some
 number. But on the other hand, one can require that the
 {\it physical} length is taken to its minimal value. But the
 physical length depends on the scale factor and so does $\bar{\mu}$,
 $\mu\to\bar{\mu}\equiv\bar{\mu}(|p|)$, $|p|=a^2$
 \cite{Ashtekar}.\footnote{The interpretation of the area operator in
   loop quantum gravity is
   still unclear. The operator itself is not a Dirac observable and
   thus even in the sense defined by the loop quantum gravity community not an
   observable (though it can become a Dirac observable when matter is
   added). In how far the area operator relates to a physical area is
   thus unsettled.} 
Depending on which of the two viewpoints is taken, one arrives at a
difference equation either in eigenvalues of the triad, $\mu$, or in
eigenvalues of the volume operator, $v$. (As discussed in the introduction, this is more suitably understood as a volume-dependent creation of vertices an thus a refined implementation of the action of the full Hamiltonian constraint.)\\
The Wheeler--DeWitt equation is recovered in the respective continuum
limit. Starting from the same factor-ordering of the difference
equation, both versions of it, the one equidistant in $\mu$, the other
equidistant in $v$, have the same Wheeler--DeWitt limit, meaning they
yield the same factor-ordering of the Wheeler--DeWitt equation. As in
\cite{Bojowald} and \cite{Ashtekar,Vandersloot} different
factor orderings have been employed, we comment briefly on both of
them here.\\ 
The question of whether the preceding result persists in loop quantum cosmology can then be reformulated as the question whether the results obtained in
Sec. \ref{Quantization} are robust with respect to a change 
of factor ordering. 

\subsection{Non-covariant factor ordering} 
\label{mu0}

The question here is under which conditions the continuum limit is justified. It is justified if the discreteness of
spacetime is negligible compared to the length scales occuring in the
model. For large scale factor, $a\gg\mu_0$, one can argue that
the limit $\mu_0\to 0$ is a sensible
approximation. Thus singularity avoidance for large-scale
singularities as, for example, the big rip or big brake, in the loop quantum cosmology framework, reduces to singularity avoidance induced by the Wheeler--DeWitt equation.

The Wheeler--DeWitt equation emerging in the continuum limit of the difference equation employed in \cite{Bojowald} is

\be
\frac{\hbar^2}{2}\left[\frac{\kappa^2}{6}a^2\frac{\partial^2\Psi}{\partial
  a^2}-\frac{\partial^2\Psi}{\partial\phi^2}\right]-a^6\frac{\tilde{V_0}}{|\phi|}\Psi=0 \ ,
\ee 
which differs from \eqref{WheelerDeWitt2} by the choice of factor-ordering.
Making the ansatz $\Psi(a,\phi)=\sum_k A(k)C_k(a)\varphi_k(a,\phi)$ and
requiring $\varphi_k(a,\phi)$ to be a solution of 

\be
\left(\frac{\hbar^2}{2}\frac{\partial^2}{\partial\phi^2}+a^6\frac{\tilde{V_0}}{|\phi|}\right)\varphi_k(a,\phi)=-E_k(a)\varphi_k(a,\phi)\ ,
\ee
one finds as before the solution

\be
\varphi_k(x_k)=N_kx_ke^{-\frac{x_k}{2}}L^{1}_{k-1}(x_k)\ ,
\ee
where $x_k=2\sqrt{-\frac{2E_k(a)}{\hbar^2}}|\phi|$ and
$E_k(a)=-\frac{1}{2\hbar^2k^2}\tilde{V_0}^2a^{12}$. Then the equation for
$C_k(a)$ is given by

\be
\frac{{\mathrm d}^2C_k(a)}{{\mathrm d}a^2}-\frac{6{\tilde{V_0}}^2}{\hbar^4k^2\kappa^2}a^{10}C_k(a)=0\ ,
\ee
which is solved by 

\bea
C_k(a)=&c_1\sqrt{a}J_{\frac1{12}}\left(\frac16\sqrt{-\frac{6{\tilde{V_0}}^2}{\hbar^4k^2\kappa^2}}a^6\right)\nonumber\\ 
+&c_2\sqrt{a}Y_{\frac1{12}}\left(\frac16\sqrt{-\frac{6{\tilde{V_0}}^2}{\hbar^4k^2\kappa^2}}a^6\right)\ .
\eea
The complete solution has an analogous form to the quantum
geometrodynamical formulation in Sec. III B. 
The decisive result is that, because only the factor ordering of the {\em gravitational part} has been changed compared to
\eqref{WheelerDeWitt2}, the solution for $\varphi_k(\phi,a)$ handles the
singularity avoidance in this framework as well. 

\subsection{Covariant factor ordering}
The factor ordering in the more recent paper
\cite{Ashtekar,Vandersloot} yields the Laplace--Beltrami
factor ordering for the Wheeler--DeWitt equation in the continuum
limit. As this is the factor ordering we employed throughout this
paper, the results of the previous sections carry over to the loop
quantum cosmology analysis without alteration. 
\\
Note, though, that a consistent loop quantization requires a polymer
representation of the matter fields as well. This would require a Bohr
compactification of $\phi$ which may bound the approximate potential
$V(\phi)=\frac{\tilde{V_0}}{\vert\phi\vert}$ from above. As the
vanishing of the wave function at $\phi=0$ is related to the
divergence of the potential at this point, it is not clear whether the
previous results would survive in the polymer representation; namely, it is
imaginable that the regularity condition and thus the ensuing
condition that $\varphi_k(\phi=0,\alpha)=0$ becomes redundant. This
has to be investigated in future publications. 

\section{Discussion and Outlook}

We studied a Friedmann--Lema\^{\i}tre model with a scalar field
obeying an 
`anti-Chaplygin' equation of state. This model classically ends with a
big-brake singularity. The singularity stands out because of its
negatively diverging second derivative of the scale factor. This works
as an infinitely strong `brake', forcing the derivative of the scale
factor to go to zero. The evolution of the scale factor stops. Upon
quantizing this model in the quantum geometrodynamical framework, we
are led to the Wheeler--DeWitt equation. It can be solved in
the vicinity of the big-brake singularity. A separation ansatz yields a
Schr\"odinger-type equation for the hydrogen atom for $\phi$ (which
here plays the role of the radius in the quantum mechanical equation).
Solutions to this equation vanish at $\phi=0$, which corresponds
to the singularity. 
Thus, independent of the choice of initial conditions, whatever
wave packet is constructed out of these solutions, it is condemned to
vanish at the singularity. 
Therefore we can conclude that in this
model as well, the large scale, soft, future singularity is removed
from the quantum theory.\\
The same model was also studied in loop quantum cosmology. Here, the
analysis was restrained to the vicinity of the big-brake
singularity. Two different factor-orderings were studied. For both we could
corroborate avoidance of the big-brake singularity.\\
Due to the special form of the potential, we were able to solve the
model in the geometrodynamical framework also in the vicinity of the
big-bang singularity. The choice of variables enforces a boundary
condition which causes the wave function to vanish at the big
bang. This singularity is thus also eliminated in the quantum
theory. The imposition of boundary conditions on both ends of the
evolution, near the big bang and near the big brake, should imply some
kind of quantization rule upon matching the wave packets in both
regimes. Such a matching has not been carried out.

What are the implications of this singularity removal?
Since the wave packet starts to spread when approaching the region
where the classical singularity would lurk, this means that the end of
the classical evolution is reached. Any information gathering and
utilizing system would stop to exist. A similar scenario may happen when
the turning point of a classically recollapsing quantum universe is
approached \cite{KieferZeh}. Classical time then comes to an end. The
details of such a scenario can, of course, only be discussed if one
goes beyond minisuperspace: the treatment of concepts such as entropy and
the arrow of time need additional degrees of freedom
\cite{OUP,Zeh}. We plan to return to this issue in a future publication.

\section*{Acknowledgements}

We thank Martin Bojowald, John Barrow,
 and Mariusz D\c{a}browski for discussions and comments. 
A.Y.K. is grateful for the hospitality of the Institute for
Theoretical Physics at the University of Cologne where part of this
work was done under the grant 436 RUS 17/8/06 of the German Science
Foundation (DFG).
B.S. thanks the
Friedrich-Ebert-Stiftung for financial support. She also thanks the
Institute for Gravitational Physics and Geometry at the Pennsylvania
State University for kind hospitality while part of this work was done.

\begin{appendix}

\section{Validity of Born--Oppenheimer approximation}

The Born--Oppenheimer approximation consists in neglecting 
cross-terms of the form

\begin{equation*}
A_{nm}=\int_0^\infty\hspace{-2mm}\d\phi\hspace{1mm}
\varphi_n(x_n)\frac{\del}{\del\alpha}\varphi_m(x_m)\ .
\end{equation*}
To be able to give some indication of the quality of the
approximation, it is necessary to evaluate these terms.
Carrying out the differentiation, one finds

\ben
\lb{crossterm}
A_{nm}=&3\int_0^\infty\d\phi\hspace{1mm}
x_nx_me^{-\frac{x_m+x_n}2}L^1_{n-1}(x_n)\\ &\left(L^1_{m}(x_m)-L^1_{m-2}(x_m)\right)\ .
\een
This integral can be evaluated using the general formula \cite{Gradshteyn}

\ben
& &\int_0^\infty\d xe^{-bx}x^aL^a_n(\lambda x)L^a_m(\mu x)\\
&=&\frac{\Gamma\left(m+n+a+1\right)}{m!n!}
\frac{\left(b-\lambda\right)^n\left(b-\mu\right)^m}{b^{m+n+a+1}}\times\\
& &F\left(-m, -n; -m-n-a;
  \frac{b\left(b-\lambda-\mu\right)}{\left(b-\lambda\right)\left(b-\mu\right)}\right)\ .
\een
Starting with the first part of the integral in \eqref{crossterm},

\begin{equation*}
X_1\equiv\int_0^\infty\d\phi\hspace{1mm}\phi^2
e^{-k\phi}L^1_{n-1}(\lambda\phi)L^1_{m}(\mu\phi)\ ,
\end{equation*}
where the short hands $k\equiv\frac{n+m}{Z(\alpha)nm}$,
$\lambda=\frac2{Z(\alpha)n}$ and $\mu=\frac2{Z(\alpha)m}$ have been
used, one finds

\ben
X_1&=\left(-1\right)^{n-1}
\frac{Z(\alpha)^3}{2}m^2n(m+1)^2(n+m-1)!(m+1)!\\
&\left(\frac{nm}{m-n}\right)^2\left(\frac{m-n}{m+n}\right)^{m+n}
\times\left[X_{11}-X_{12}-X_{13}\right]\ .
\een
Here,

\ben
X_{11}&=&2\left(\frac{m-n}{m+n}\right)F\left(-m,-n+1;-m-n;\left(\frac{m+n}{m-n}\right)^2\right)\\
X_{12}&=&F\left(-m+1,-n+1;-m-n+1;\left(\frac{m+n}{m-n}\right)^2\right)\\
X_{13}&=&\frac{(n+m+1)}{m(m+1)}\frac{(m-n)^2}{(m+n)}\times\\
& &F\left(-m-1,-n+1;-m-n-1;\left(\frac{m+n}{m-n}\right)^2\right)\ .
\een
Similarly, the second part,

\begin{equation*}
X_2\equiv\int_0^\infty\d\phi\hspace{1mm}\phi^2
e^{-k\phi}L^1_{n-1}(\lambda\phi)L^1_{m-2}(\mu\phi)\ ,
\end{equation*}
can be integrated to

\ben
X_2&=&\left(-1\right)^{n-1}
\frac{Z(\alpha)^3}{2}mn(m-1)^2(m-2)(m+n-3)!\\
& &\left(\frac{nm}{m-n}\right)^2\left(\frac{m-n}{m+n}\right)^{m+n}\\
& &\times\left[X_{21}-X_{22}-X_{23}\right]\ ,
\een
where

\ben
X_{21}&=&2\frac{(m+n-2)}{(m-2)}\left(\frac{m-n}{m+n}\right)\\
& &F\left(-m+2,-n+1;-m-n+2;\left(\frac{m+n}{m-n}\right)^2\right)\\
X_{22}&=&F\left(-m+3,-n+1;-m-n+3;\left(\frac{m+n}{m-n}\right)^2\right)\\
X_{23}&=&\frac{(m+n-2)(m+n-1)}{(m-2)(m-1)}\left(\frac{m-n}{m+n}\right)^2\\
& &F\left(-m+1,-n+1;-m-n+1;\left(\frac{m+n}{m-n}\right)^2\right)\ .
\een
Taking into account the prefactors, one finds

\begin{equation*}
A_{nm}=Z(\alpha)3m\left(\frac2n\right)\left(\frac2m\right)
\left[\frac{X_1}{Z(\alpha)^3}-\frac{X_2}{Z(\alpha)^3}\right]\ .
\end{equation*}
\\
So $A_{nm}\propto Z(\alpha)\propto a^{-6}$. For $a=a_{\star}$ it thus
takes its minimal value. This shows that the Born--Oppenheimer
approximation is fulfilled best when one approaches the region of the
classical singularity. 

\end{appendix}
 

\end{document}